\documentclass{appolb}
\usepackage{graphicx}
\usepackage{doi}
\usepackage{amsmath}
\graphicspath{{figs/}}

\newcommand{\dF}{{^{^*}\!\!F}}

\newcommand{\sB}{{\mathcal{B}}}
\newcommand{\sR}{{\mathcal{R}}}
\newcommand{\sQ}{{\mathcal{Q}}}

\begin{document}

\title{Neutrino cooled disk in post-merger system studied via numerical
  GR MHD simulation with a composition-dependent equation of state
  \thanks{Presented at the $8^{\rm th}$ Conference of the Polish Society on Relativity, Warsaw,
    Poland, 19-23 September, 2022.}%
}
\author{Agnieszka Janiuk
\address{Center for Theoretical Physics of the Polish Academy of Sciences, \\ Al. Lotnik\'ow 32/46, 02-668 Warsaw, Poland}
\\
%{Third Author of different affiliation
%}
%the Name(s) of other Author(s)
%\address{affiliation}
}
\maketitle
\begin{abstract}
  The code HARM\_COOL, a conservative scheme for relativistic magnetohydrodynamics,
is being developed in our group and works with a tabulated equation of state of dense matter. This EOS can be chosen and used during dynamical simulation,
instead of the simple ideal gas one.
In this case, the inversion scheme between the conserved and primitive variables
is not a trivial task. In principle,
the code needs to solve numerically five coupled
non-linear equations at every time-step.
The 5-D recovery schemes were originally implemented in HARM and worked accurately for a simple polytropic EOS which has an analytic form.
Our current simulations support the composition-dependent EOS, formulated in terms of rest-mass density, temperature and electron fraction.
In this proceeding, I discuss and compare several recovery schemes that have been included in our code. I also present and discuss their convergence tests.
Finally, I show set of preliminary results of a numerical simulation, addressed to the post-merger system formed after the binary neutron stars (BNS) coalescence.
\end{abstract}

%\PACS{PACS numbers come here}

\section{Introduction}

The first detection of a gravitational wave signal accompanied by an electromagnetic counterpart was announced on August 17, 2018, by the LIGO-Virgo team \cite{abbott}.
The signal originated from the coalescence of two neutron stars of masses in the range of 1.17-1.60 $M_{\odot}$ and the total mass of the system of 2.74 $M_{\odot}$. It was followed by a short weak gamma ray burst, observed 1.7 seconds after the GW signal.
Hence, the theoretical prediction that this
class of short gamma ray bursts originate from compact binary mergers has been proven by the detection of the source GW 170817.

It was also suggested before that the radioactivities from dynamical ejecta after the first neutron star had been disrupted, can
power an electromagnetic signal \cite{lipaczynski1998}.
%(e.g. Li & Paczynski 1998; Tanvir et al. 2013; Korobkin et al. 2012)
Subsequent accretion onto a newly formed black hole
can provide bluer emission, if it is not absorbed by precedent ejecta
\cite{SiegelMetzger2017, korobkin}.
%(Tanaka M., 2016, Berger 2016, Siegel & Metzger 2017)
In this case, a day-timescale emission comes at optical wavelengths from lanthanide-free components of the ejecta, and is followed by a week-long emission with a spectral peak in the near-infrared (NIR).
This two component model fits well with observations of the kilonova, detected in coincidence with the source GRB-GW 170817.

In our studies, we investigate the very last stage of the system, namely the post-merger black hole accretion disk.
Due to its high density, the accretion disk in post-merger system is opaque to photons. Neutrinos are produced via $\beta$-reactions, electron-positron pair annihilation, and plasmon decay and provide cooling mechanism.
The plasma is composed of free nucleons, pairs, and Helium. In the outer regions, heavier nuclei can be also synthesized, under conditions of nuclear statistical equilibrium (NSE, see \cite{Janiuk2014} for details).
The kilonova signal is produced in the equatorial wind
outflows, launched from the disk via magnetic instabilities.
These ejecta are dominated by thermal energy of the dense plasma, and are accelerated to mildly relativistic velocities
(of about 0.2-0.3 $c$, \cite{Janiuk2019}). The matter is highly neutronized there, so due to the r-process nucleosynthesis, copious amounts of unstable heavy isotopes are formed in these winds, and power the Infrared/Optical emission through their radioactive decay.

\section{Numerical modeling}

Our study of the winds launched from the accretion disk, is done by evolving the general relativistic magnetohydrodynamic (GR MHD) equations in time. We use the HARM (High Accuracy Relativistic Magnetohydrodynamics) code \cite{Gammie2003}  which is a conservative and shock capturing scheme. The numerical scheme advances the conserved quantities from one time step to the next by solving a set of non-linear hyperbolic equations
for continuity, energy-momentum conservation and magnetic induction.
In the GR MHD scheme they read:

\begin{equation}
    \nabla_\mu(\rho u^{\mu})= 0, \\
    \nabla_\mu(T^{\mu\nu}) = 0 , \\
    \nabla_\mu(u^\nu b^\mu - u^\mu b^\nu) = 0
\end{equation}
where
\begin{equation}
     T^{\mu\nu} = T^{\mu\nu}_{\rm gas} + T^{\mu\nu}_{\rm EM}
\end{equation}
is contributed by
\begin{equation}
  T^{\mu\nu}_{\rm gas} = \rho h u^\mu u^\nu + p g^{\mu\nu} = (\rho+u+p)u^{\mu}u^{\nu}+pg^{\mu\nu},
\end{equation}
\begin{equation}
    T^{\mu\nu}_{\rm EM} = b^2 u^{\mu} u^{\nu}+ \frac{1}{2} b^2 g^{\mu \nu} - b^{\mu}b^{\nu}, ~~~
    b^{\mu} = u^{*}_{\nu} F^{\mu \nu}.  
\end{equation}

In the stress-energy tensor composed of the gas and electromagnetic terms, $u^{\mu}$ is the four-velocity of the gas, $u$ is the internal energy, $\rho$ is the density, $p$ is the pressure, and $b^{\mu}$ is the magnetic four vector. $F$ is the Faraday tensor and in a force-free approximation, we have $E_{\nu} = u^{\nu}F^{\mu \nu} = 0$. The unit convention is adopted such that $G=c=M=1$.

Our initial conditions mimic the configuration after the transient (hyper-massive-neutron star, HMNS) object has collapsed to a black hole, and
a pressure equilibrium torus \cite{FM1976} has been formed.
In this solution, the angular momentum along the radius of the disk is constant.

We parameterize our models with black hole Kerr parameter, $a$, and
the inner radius and radius of pressure maximum, $r_{\rm in}$, and $r_{\rm max}$, respectively.
The current simulations are run in an axisymmetric setup,
i.e. 2D, with a resolution 256x256 points in the $r$, and $\theta$ directions.
The numerical code works in Kerr-Schild coordinates, which enables the matter to smoothly accrete under the horizon.

The torus is embedded in an initially poloidal magnetic field, prescribed with the vector potential of
$(0,0,A_{\varphi})$, with
%\begin{equation}
$A_{\varphi}=({\bar{\rho} \over \rho_{\rm max}} - \rho_{0})$ %\times r^{5}
%\end{equation}
where we use offset of $\rho_{0}=0.2$.

We parameterize the field strength with plasma $\beta$ defined as the ratio of the
gas-to-magnetic pressure, $\beta = p_{\rm gas} /p_{\rm mag}$.
Here $p_{\rm gas} = (\gamma - 1)u_{\rm max}$ and $p_{\rm mag} = b_{\rm max}^2/2$, where $u_{\rm max}$ is the internal energy of the gas at the radius of maximum pressure.

\section{Chemical composition and structure of the disk}

On eof significant chalelnges for numerical simulation is neutrino treatment.
Neutrinos carry away energy and lepton number, so they alter electron fraction and composition of ejected material. 
Dynamical simulations must consider the realistic equation of state (EOS) and impact of neutrinos in the optically thin and thick regions.
Lepton number conservation is expressed as follows:

\begin{equation}
\nabla_{\mu}(n_{e} u^{\mu}) = \sR/m_{b}; ~~ m_{b}=\rho/n_{b} ; ~~ Y_{e} = {n_{e} \over n_{b}} = { m_{b} n_{e} \over \rho}
\end{equation}

%~~ \nabla_{\mu}T^{\mu}_{\nu} = 0; ~~ \nabla_{\mu}(u^{\nu}b^{\mu}-u^{\mu}b^{\nu}) = 0 \]
where $\sR$ is the net rate of neutrino and antineutrino volume number densities in the fluid frame, and $Y_{e}$ is the electron fraction.

Since the baryons dominate the rest-mass density, the baryon number conservation equation turns to the regular continuity equation.
In the energy-momentum conservation equation, we must introduce an additional source term due to heating and cooling by neutrinos:
\begin{equation}
  \nabla_{\mu}T^{\mu}_{\nu} = \sQ u_{\nu},
\end{equation}
where $\sQ$ is the energy change per unit volume due to neutrino emission.

\subsection{Conserved and primitive variables}

Conservation equations solved by the GR MHD code can be expressed in a flux conservative form \cite{Siegel2018}, and explicit choice of the conserved
variables (which are analytic functions of primitive ones) is to some extent arbitrary.
The recovery schemes for the primitive variables need numerical root finding, and can be broadly devided into two categories.
%\begin{itemize}
%\item
  In a 2D scheme, there are two independent variables, e.g, $v^{2}$ and specific enthalpy $W$.
  Temperature is obtained from the EOS
  tables, solving $h=h(\rho, T, Y_{e})$, and pressure and temperature are also obtained by Newton-Raphson method for $W$ and $v^{2}$.
%\item
Alternatively, the system of GR MHD equations is reduced to 3 equations, which have three unknowns.
The chosen independent variables can be:
$\gamma$, $T$, and $W=h\rho \gamma^{2}$.
Pressure is interpolated from EOS tables, as $P(\rho, T, Y_{e})$.
%\end{itemize}

\subsection{Recovery transformation}

In the 2D recovery scheme of \cite{Noble2006}
%Noble et al. (2006),
the dimensionality of the recovery
problem is reduced by making use of certain scalar quantities that can be
computed from the conservatives.
To avoid numerical pathologies o the $1D_{W}$ scheme near the roots,
this 2D scheme solves simultaneously the set of two equations:
\[ f_{1}: ~~ \tilde Q^{2}= v^{2} (\sB^{2} + W)^{2} - {{(Q_{\mu} \sB^{\mu})^{2} (\sB^{2} + 2W) } \over{ W^{2} }} \]
\[ f_{2}: ~~ Q_{\mu} n^{\mu} = - {\sB^{2} \over 2} (1 + v^{2}) + {{(Q_{\mu} \sB^{\mu})^{2}} \over {2 W^{2}}} - W + p(u, \rho) \]

The independent variables used in this scheme are defined as:
$ Q_{\mu} = -n_{\mu}T^{\nu}_{\mu} = \alpha T^{t}_{\mu} $;
where $\tilde Q^{\nu} = j^{\nu}_{\mu} Q^{\mu}$
is energy-momentum density in the normal obsever frame, and 
$ D= -\rho n_{\mu} u^{\mu} = \alpha \rho u^{t} = \gamma \rho $;
is mass density in the observer's frame,
and 
$\sB^{i}= \alpha B^{i} = \alpha \dF^{it} $
is  magnetic 3-vector.
Here, 
$ w = \rho+u+p$ ~~with  $W = w \gamma^{2}$ is  enthalpy.
%(note it is original notation).
To solve this set of equations
by means of Newton-Raphson method, the Jacobian matrix with $\partial f_{1} \over \partial(v^{2})$, $\partial f_{2} \over \partial(v^{2})$, $\partial f_{1} \over \partial W$, and $\partial f_{2} \over \partial W$ is needed.
Note that this scheme does not require an analytic EOS, and derivatives of pressure wtr. to $\rho$, $v^{2}$, and $u$ may be computed from tables using finite difference method.

In the 3D recovery scheme of \cite{Siegel2018}, the system is extended to solve 3 equations, on $W, z$, and $T$, by adding a constraint on the internal energy given by EOS tables:
\[ f_{1}: ~~ [\tau + D -z - B^{2} + {B^{i}S_{i} \over 2 z^{2}} + \rho ] W^{2} - {B^{2} \over 2} = 0 \],
\[ f_{2}: ~~ [(z+B^{2})^{2} - S^{2} - {{2z +B^{2}} \over {z^{2}}} (B^{i}S_{i})^{2}] W^{2} - (z+B^{2})^{2} = 0 \],
\[ f_{3}: ~~ \epsilon - \epsilon(\rho, T, Y_{e}) = 0 \].
Here the temperature is employed directly as an unknown through
$ \epsilon(W, z T) = h -1 - {P \over \rho} = {z-DW - \rho W^{2} \over DW }$
and does not require inversion of the EOS.
Notice here different notation: $S_{i}$ is energy-momentum density, $W$ is Lorentz factor, $z$ is enthalpy, and $\tau = -(n_{\mu}n_{\nu} T^{\mu \nu} +D)$.

 \begin{figure}
   \centering
   \includegraphics[width=0.48\textwidth]{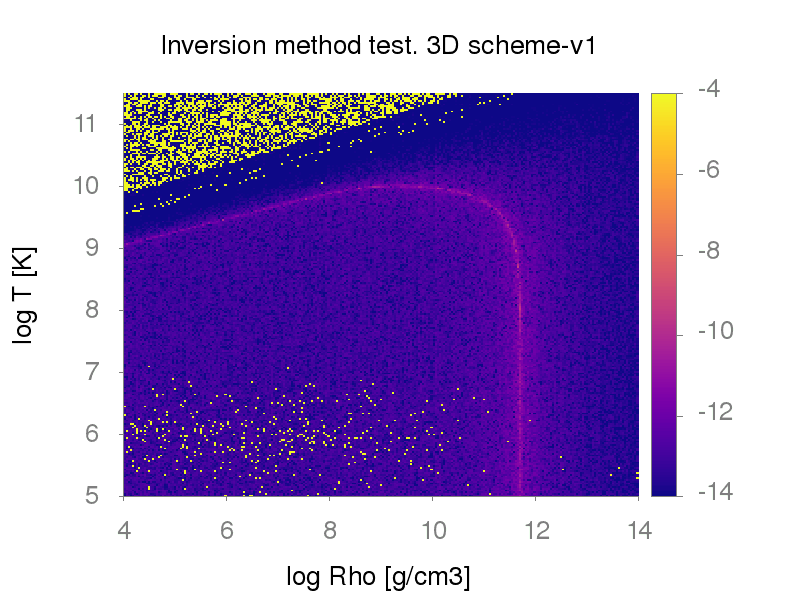}
\includegraphics[width=0.48\textwidth]{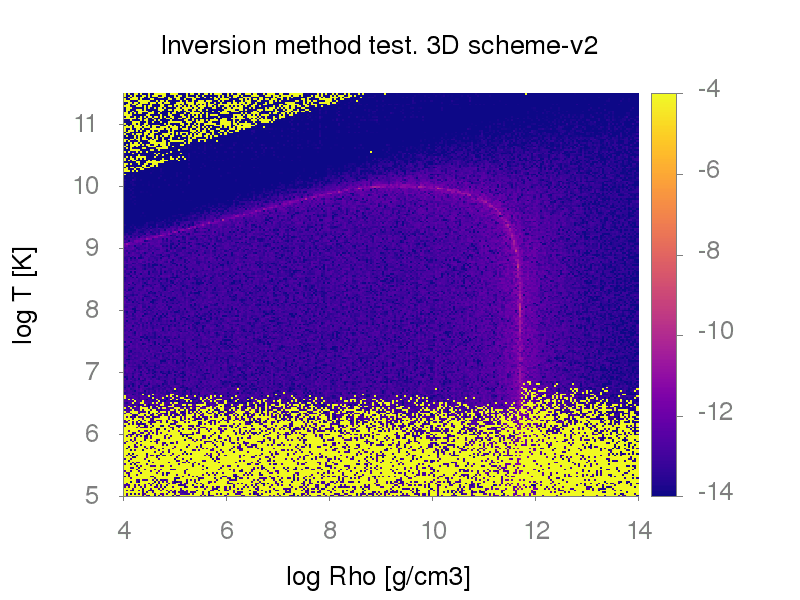}
\includegraphics[width=0.48\textwidth]{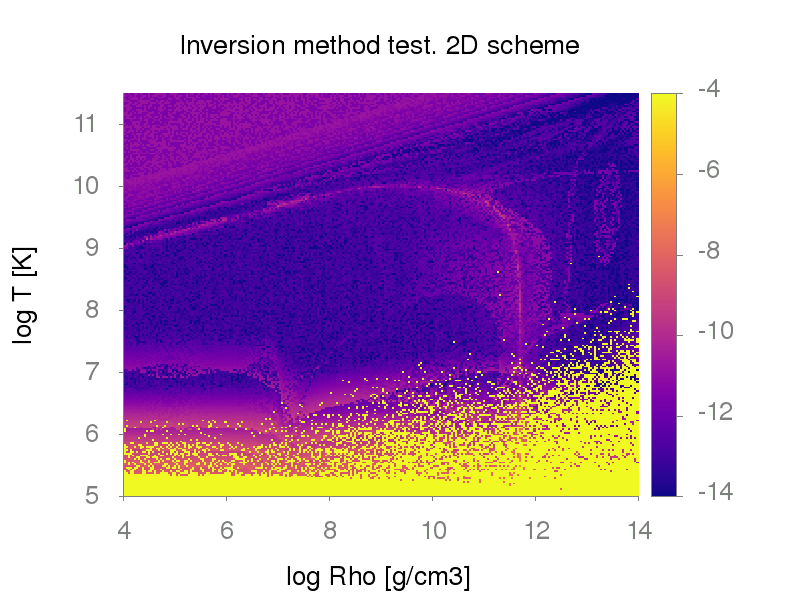}
\includegraphics[width=0.48\textwidth]{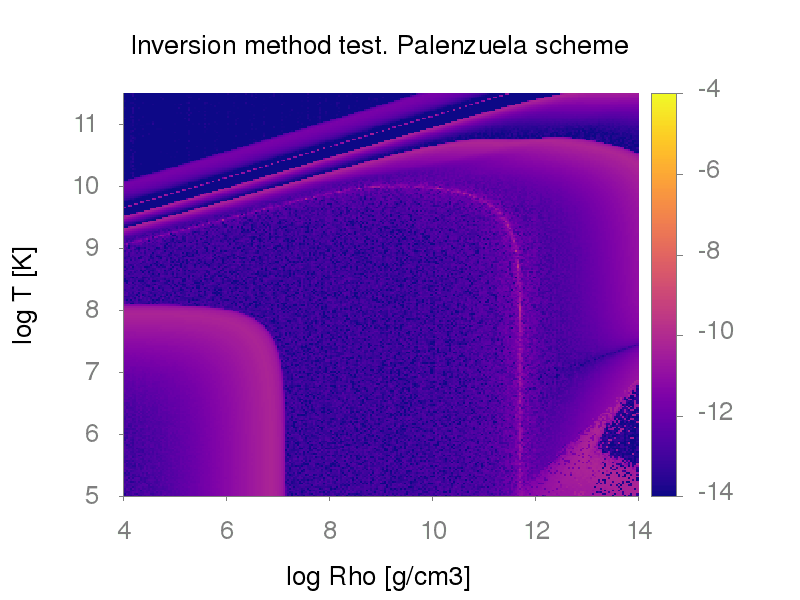}
   \caption{Convergence test results for the 3D method (2 versions,
     with (1): we compute specific internal energy from state vector x and conservatives
  as in Eq. (25) in \cite{Cerda2008}, and solve $f_{3}$
  or (2): we compute pressure from state vector x and conservatives and solve $f_{3}$ ), shown in the top panels, the 2D method (bottom panel, left) and
  Palenzuela method (bottom panel, right).
      }
   \label{fig:conv_test}
   \end{figure}

 In method proposed in \cite{Palenzuela2015}
 the scheme is solving 1D equation, for the rescaled variable, $ \chi = {{\rho h \gamma^{2}}\over {\rho \gamma} }$.
Other quantities are also rescaled, accordingly, to give Lorentz factor,
and we give the brackets for $\chi$: $ 2 - 2 {{Q_{\mu}n^{\mu} +D}\over {D}} - {\sB^{2}\over D} < \chi <  1  - {{Q_{\mu}n^{\mu} +D}\over {D}} - {\sB^{2}\over D} $.
The equation
\[ f(\chi) = \chi - \tilde\gamma (1+\tilde\epsilon + {\tilde P \over \tilde\rho}) = 0 \]
is solved, with $\tilde{P} = P(\tilde\rho, \tilde\epsilon, Y_{e})$ found in tables.

The last method has proven to be most robust, for our wide parameter space. It works with smaller errors, while the performance spped is also slower.
We performed the convergence tests to explore which recovery transformation works best for the wide parameter space.
The parameters used for testing routines were: $Y_{e}=0.1$, $\gamma=2$, $p_{\rm gas}/p_{\rm mag}=10^{5}$.
We derived the conserved variables in Kerr metric, and then computed primitives perturbed by a factor of 1.05.
The variables recovered through the 2D scheme were compared to the unperturbed, to calculate the total error, summed over all primitives
${\rm Err} = \Sigma_{k=0, NPR} (P_{k} - \bar{P_{k}})^{2}$.
Figure \ref{fig:conv_test} shows the results for all 3 methods probed.

\subsection{Neutrino transport}

We employ the neutrino leakage scheme that computes a gray optical depth estimate along radial rays for electron neutrinos, electron antineutrinos, and heavy-lepton neutrinos (nux), and then computes local energy and lepton number loss terms.
The source code of the scheme has been publicly available and
downloaded from \textit{https://stellarcollapse.org}. Details are described e.g. in \cite{Ott2012}.
%( O'Connor \& Ott (2010), Ott et al. (2012)).
The initial tests were done within an optically thin regime for neutrinos.
\begin{equation}
\tau(r,\theta,\phi) = \int_{r}^{R} \sqrt{\gamma_{rr}}\bar\kappa_{\nu_{i}} dr'< 2/3.
\end{equation}

  \section{Results}

\begin{figure}[htb]
%\centerline
\centering
{%
\includegraphics[width=0.3\textwidth]{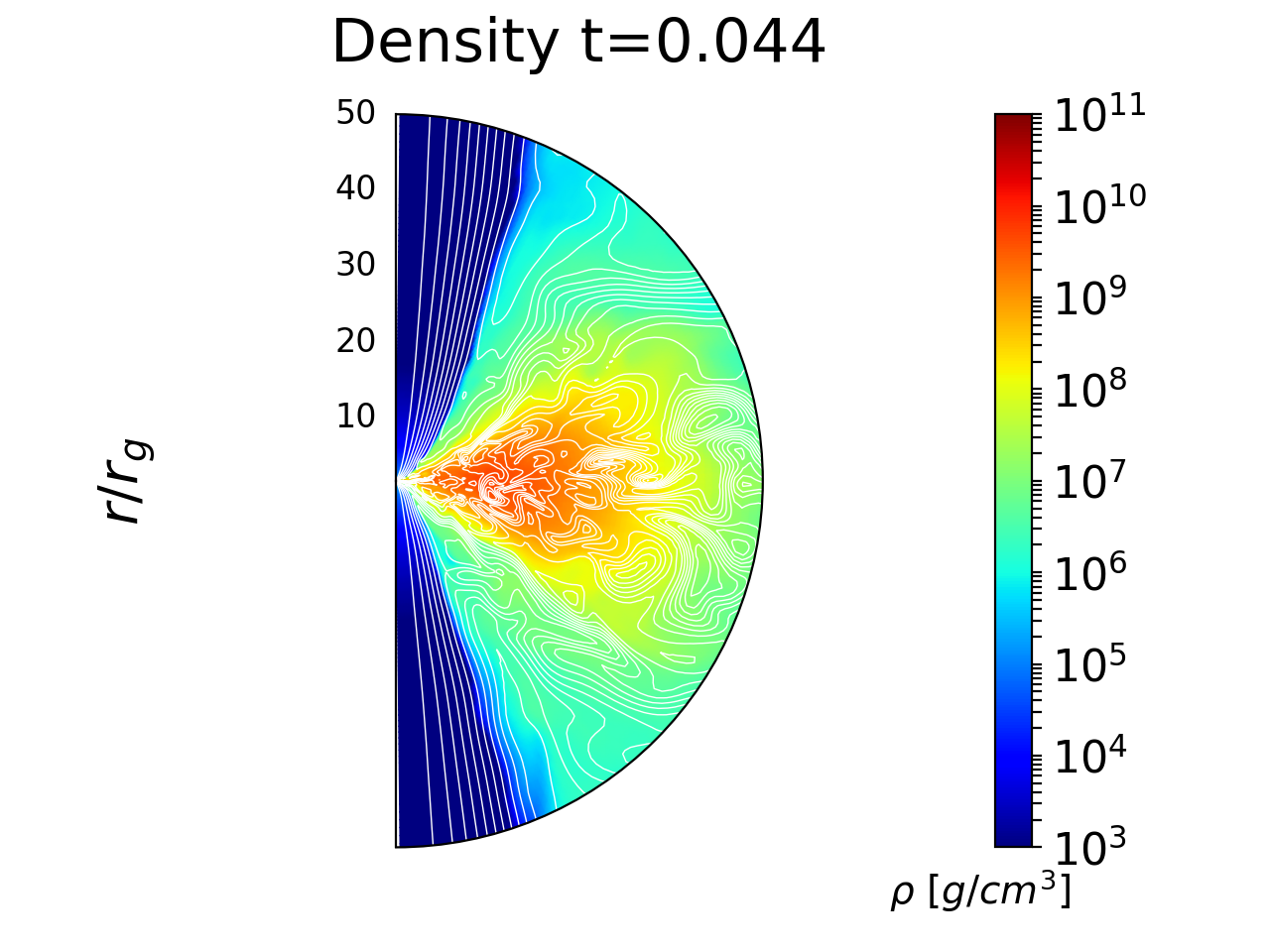}
\includegraphics[width=0.3\textwidth]{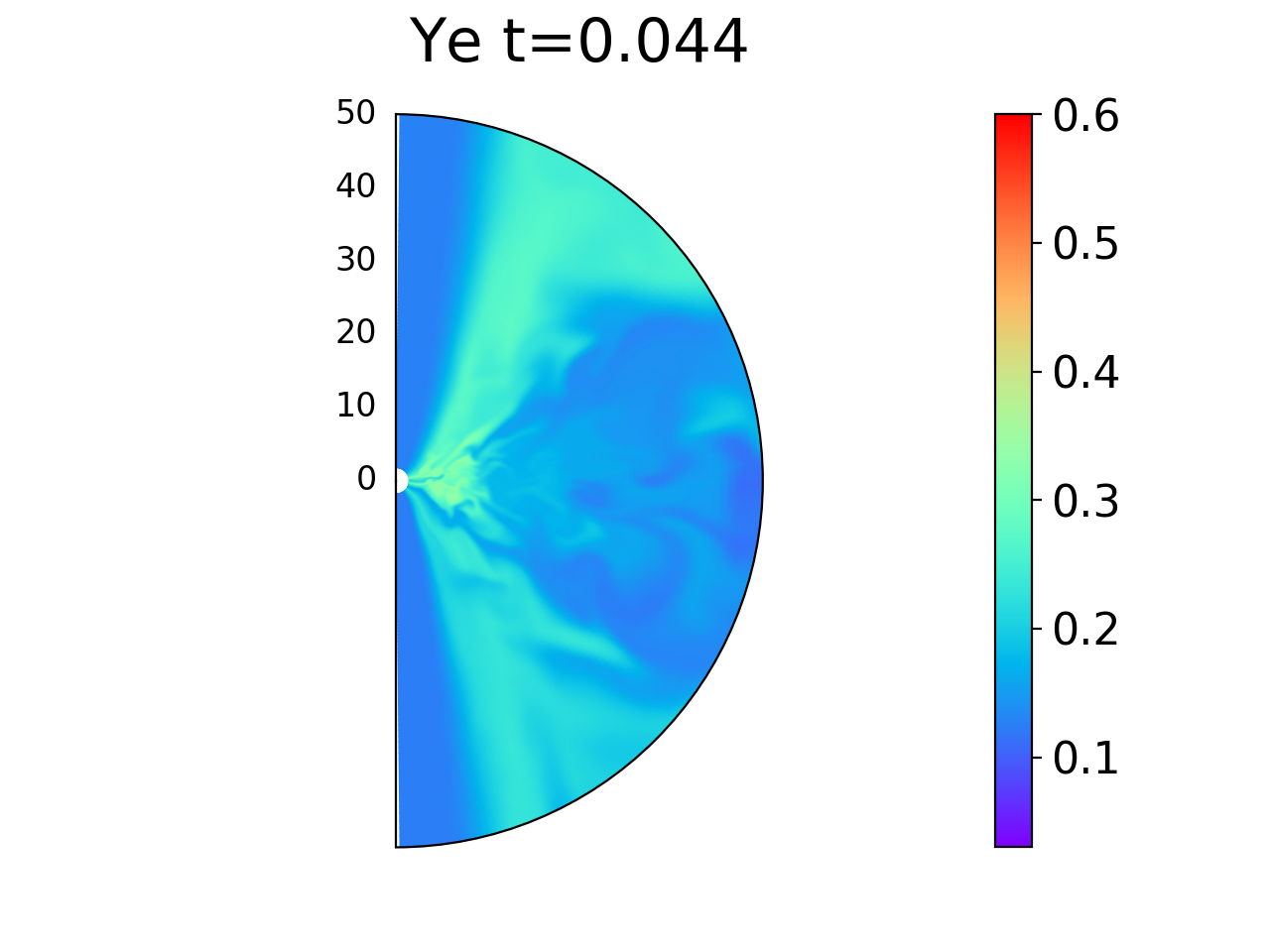}
\includegraphics[width=0.3\textwidth]{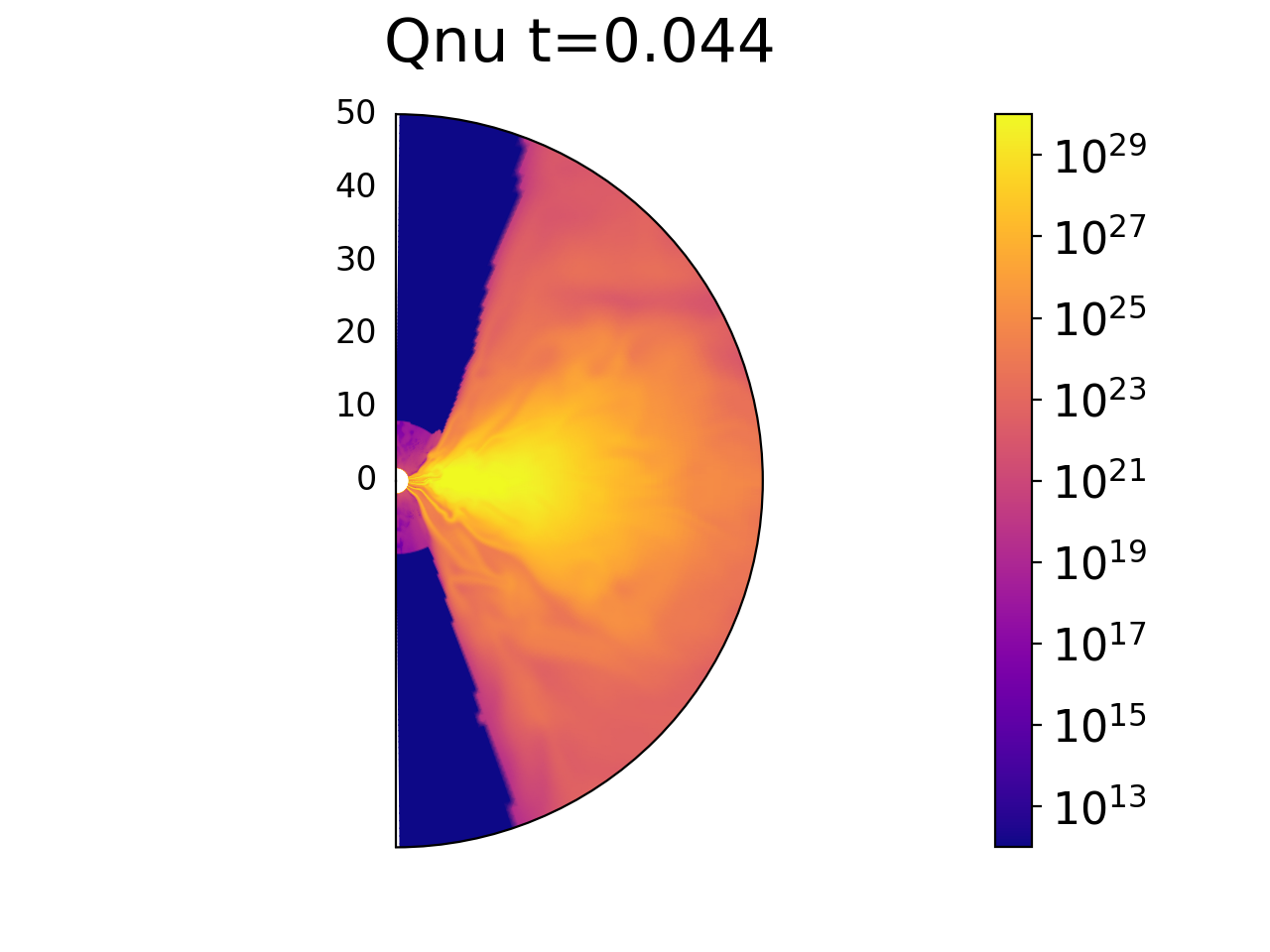}

}
\caption{Maps of density and magnetic field contours (left), electron fraction (middle) and neutrino emissivity (right panel). Model assumes optically thin plasma for neutrinos. Inversion scheme used the Palenzuela method.
  The model parameters are disk mass = 0.13 $M_{\odot}$, 
  gas-to-magnetic pressure ratio
  $\beta=50$ and black hole spin $a=0.6$.
  The color maps are in logarithmic scale and are taken at $t=0.044$ s. 
   }
\label{fig:model_opt_thin}
\end{figure}

The structure of the flow in the evolved state is shown in Fig. \ref{fig:model_opt_thin}. We present polar slices, taken at the evolved time of the simulation ($t=44 ~ms$, which is equivalent to the 3000 gometric time units, $t_{g}$, for black hole mass of 3 $M_{\odot}$). Parameters of this model are black hole spin $a=0.6$, initial torus radius $r_{\rm in}=4 r_{g}$ and pressure maximum radius of $r_{\rm max}=11.8 r_{g}$.
We also assume initial electron fraction in the torus of $Y_{\rm e, disk}=0.1$, and in the atmosphere it is  $Y_{\rm e, atm}=0.45$. The specific entropy per baryon in the torus is assumed $S=10 k_{\rm B}$. This, with the enthalpy value resulting from FM torus sulution, gives the physical density scaling: units in the code are normalized such that $\rho_{\rm max}=1$.
The torus mass in cgs units is about $0.03 M_{\odot}$.

As shown in the figure, the magnetic turbulence developed in the torus, and helped launching winds from its surface. The neutronized material is redistributed, and
electron fraction in the winds became  larger han in the torus, about $Y_{e}=0.3$.
The neutrino emissivity in the winds is couple of orders of magnitude less than in the torus.
As can be seen in the 2D map, the accreting torus has a very high neutrino
luminosity, but here neutrinos are partially trapped. A moderately high neutrino luminosity can be
observed from the wind, where the neutrino-antineutrino pairs can contribute to the heating of the plasma.

\begin{figure}
   \centering
   \includegraphics[width=0.95\textwidth]{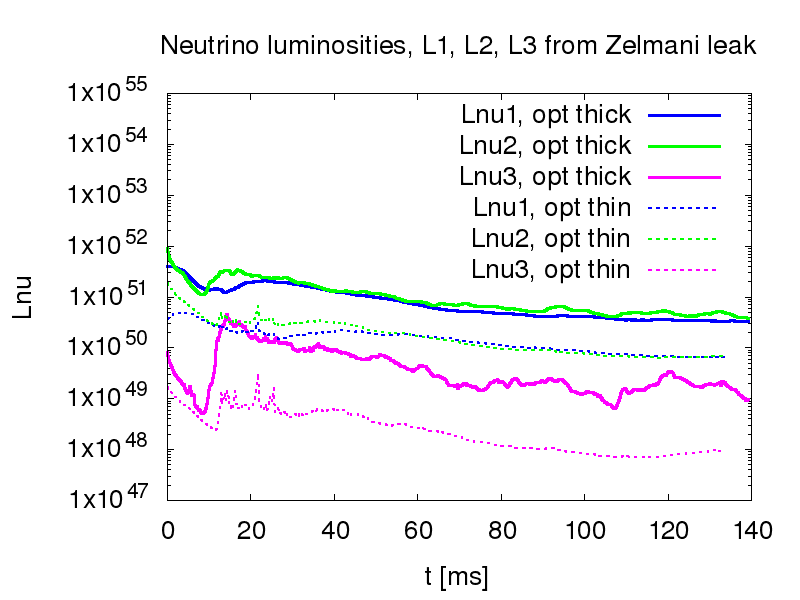}
   \caption{The electron neutrino and antineutrino (Lnu1 and Lnu2, respectively; Lnu3 denotes luminosity in other flavors)
     luminosty as a function of time, for our two models, optically thin and optically thick, calculated with the neutrino-leakage scheme. 
      }
     \label{fig:lnu}
  \end{figure}

Neutronisation of the plasma in the tours is still significant, and electron fraction values are below $Y_{e}=0.1-0.3$ in some densest regions (cf. \cite{Hossein2023}).
The neutrinos which are created in weak interactions, and
the electron-positron pair annihilation, nucleon bremsstrahlung and plasmon decay, provide an efficient cooling mechanism.
In the currently developed new numerical scheme, we substituted a simplified
description of the neutrino cooling rate given by the two-stream approximation previously used by \cite{Janiuk2019}
with a more advanced neutrino leakage scheme.
The neutrino emissivity distribution image taken at an evolved state of the torus is presented in Fig. \ref{fig:model_opt_thin} for the optically thin case. For optically thick simulation (which assumed different initial parameters as for the specific entropy), the quantitative difference can be noticed.
The neutrino and antineutrino luminosities, as a function of time, are shown in
Figure \ref{fig:lnu}.

The velocity of the wind outflows, $v \sim (0.11- 0.23)~c$,
and mass loss via unbound outflows of 2- 17\% of the initial disk mass were determined previously by \cite{Janiuk2019}.
We showed that the details are sensitive to engine parameters: BH spin and magnetisation of the disk, namely the 
more magnetized disks produce faster outflows.
We also found that the accretion disk ejecta produce heavy elements up to mass number $A\sim 200$, including platinium and gold isotopes (see details in \cite{Hossein2023}).

  %%%MOVED fom Sec 3.

  \section{Conclusions}
  
  Numerical GR MHD simulations have been widely used to model post-merger systems and
  engines of short gamma ray bursts.
We implemented a new scheme in our HARM-COOL code
to the case study of post-merger system and kilonova source.
The calculations are complex and need to cover physics of dense nuclear matter. Their performance is sensitive to the chosen recovery schemes. We tested several of them and chose the most robust for a lare parameter space in densities, temperatures, and electron fraction values, to work with the 3-parameter EOS implemented in hydrodynamical simulation.
 Proper source terms have been added in the system of equations for the neutrino losses, coupled with composition changes. We also use an advanced neutrino-leakage scheme and calculate emissivities of neutrino and antineutrino as a function of time.

 The unbound outflows, i.e. winds, are powered by both neutrinos and magnetically driven acceleration. Therefore, winds may be more dense and powerful, if neutrino heating supports the magnetically-driven wind.

\section*{Acknowledgment}
This work was supported by grant 2019/35/B/ST9/04000 from the Polish National Science Center. We used computational resources of the ICM of the Warsaw University, and the PL-Grid, under grant \textit{plggrb5}.

\bibliographystyle{plain}

\bibliography{potor8}

\end{document}